\documentclass[oldversion,rnote]{aa}
\usepackage{graphicx}
\usepackage{txfonts}

\begin{document}

\title{Jeans instability of a galactic disk 
 embedded in a live dark halo}

\author{O. Esquivel\thanks{IMPRS Fellow} 
          \and B. Fuchs}

\offprints{O. Esquivel,
\email{esquivel@ari.uni-heidelberg.de}}

\institute{Astronomisches Rechen-Institut am Zentrum f\"ur Astronomie der 
Universit\"at Heidelberg, M\"onchhofstr.~12 - 14, 69120 Heidelberg, Germany}

\date{Received  2007}

\abstract{We investigate the Jeans instability of a galactic disk embedded in 
a dynamically responsive dark halo. It is shown that the disk-halo system 
becomes nominally Jeans unstable. On small scales the instability is 
suppressed, if the Toomre stability index $Q_{\rm T}$ is higher than a certain 
threshold, but on large scales the Jeans instability sets invariably in. 
However, using a simple self-consistent disk-halo model it is demonstrated 
that this occurs on scales which are much larger than the system 
so that this is indeed only a nominal effect. From a practical point of view 
the Jeans instability of galactic disks is not affected by a live dark halo.}


\maketitle

\section{Introduction}

The classical paradigm is that at given total mass of the system a galactic 
disk is stabilized against local dynamical Jeans instabilities, if it is 
embedded in a dark halo. This can be seen for instance from Toomre's 
$Q_{\rm T}$ stability index for a stellar disk,
\begin{equation}
Q_{\rm T} = \frac{\kappa \sigma_{\rm d}}{3.36\, G\Sigma_{\rm d}}\,,
\label{eq1}
\end{equation}
where $\kappa$ denotes the epicyclic frequency of the orbits of the stars, 
$\sigma_{\rm d}$ is the radial velocity dispersion of the stars, and 
$\Sigma_{\rm d}$ the surface density of the disk. $G$ denotes the constant of 
gravitation. If all other parameters are kept constant, but 
$\Sigma_{\rm d}$ is lowered, $Q_{\rm T}$ rises and the disk becomes more stable
against Jeans instabilities. The physical reasoning is that the self-gravity 
of the disk, which has a destabilizing effect, is reduced by the surrounding 
halo. Similarly the onset of non-axisymmetric coherent large-scale 
instabilities of the entire disk such as the bar instability was thought to be
damped by a surrounding halo. Ostriker \& Peebles (\cite{OP73}) showed in
their classical numerical simulations of the dynamical evolution of a 
self-gravitating disk that the bar instability could be suppressed, if the 
disk was embedded in a halo potential. However, modern high-resolution 
simulations in which the surrounding halo is treated as a dynamically
responsive system, have shown that actually the opposite is true. 
Athanassoula (\cite{A02}) showed that in her simulations of the 
bar instability the bar grows stronger, if the disk is embedded in a live 
dark halo rather than in a static halo potential. This was explained there and,
particularly, in Athanassoula (\cite{A03}) as due to the effect of the live
halo on the angular momentum exchange within the galaxy. First doubts about 
an entirely passive role of the halo were already raised by Toomre (\cite{T77}).
These findings were supported by theoretical studies of the swing amplification
of shearing spiral density waves which is also enhanced, if the disk is
embedded in a live dark halo instead of a static potential (Fuchs \cite{F04},
Fuchs \& Athanassoula \cite{FA05}).

In this {\em note} we return to the severe local Jeans instability of a 
self-gravitating disk and investigate the effect of the presence of a live dark
halo. In the next section we demonstrate how Toomre's concept of the
$Q_{\rm T}$ parameter can be generalized in order to take into account the
effect of such a halo. In the final section we discuss implications for
realistic disk-halo systems.

\section{Modification of the $Q_{\rm T}$ stability index}

We study the Jeans instability of an infinitesimally thin galactic disk 
using the model of a patch of the galactic disk developed by 
Toomre (\cite{T64}), Goldreich \& Lynden-Bell (\cite{GL65}) and 
Julian \& Toomre (\cite{JT66}) (cf.~also Fuchs \cite{F01}). The patch is 
assumed to rotate around the galactic center and the differential rotation of
the stars is approximated as a linear shear flow. The surface density is 
assumed to be constant over the patch. Polar coordinates are approximated by 
pseudo Cartesian coordinates $(x,y)$ with $x$ pointing in the radial direction
and $y$ in the direction of rotation, respectively. Toomre (\cite{T64}) has
calculated the dynamical response of the disk to a small `ring-like'
perturbation of the gravitational disk potential of the form
\begin{equation}
\Phi_{\rm k} {\rm exp}i(\omega t+kx) 
\label{eq2a}
\end{equation}
by solving the linearized Boltzmann equation. The induced density perturbation 
can be written in the limit $\omega \rightarrow 0$ as
\begin{equation}
\Sigma_{\rm k}e^{ikx}=-\frac{\Sigma_{\rm d}}{\sigma^2_{\rm d}}
\left[ 1 - {\rm exp}\left(-\frac{k^2 \sigma_{\rm d}^2}{\kappa^2}\right)\cdot
{\rm I}_0\left(\frac{k^2 \sigma_{\rm d}^2}{\kappa^2}\right) \right] 
\Phi_{\rm k} e^{ikx} ,
\label{eq2}
\end{equation}
where I$_0$ denotes the modified Bessel function and $\Sigma_{\rm d}$ the 
background surface density of the disk, respectively. In deriving 
eq.~(\ref{eq2}) a Gaussian velocity distribution of the stars with a velocity
dispersion $\sigma_{\rm d}$ has been adopted.
The disk is assumed to be self-gravitating, so that the density -- potential
pair has to fulfill the Poisson equation implying
\begin{equation}
\Phi_k = - \frac{2\pi G}{\left| k\right|}\Sigma_k.
\label{eq3}
\end{equation}
Equations (\ref{eq2}) and (\ref{eq3}) define together a line in a space spanned 
by $Q_{\rm T}$ and the wavelength of the perturbation 
$\lambda = 2\pi /k$ expressed in units of $\lambda_{\rm crit} = 
4\pi^2G\Sigma_d/\kappa^2$, which separates neutrally stable 
($\omega^2 \geq 0$) from exponentially unstable ($\omega^2 < 0$) 
perturbations of the disk. The criterion that ensures that all perturbations
are neutrally stable is the famous Toomre criterion
\begin{equation} 
Q_{\rm T} \geq 1 \,.
\label{eq3a}
\end{equation}
The model of a local patch of a galactic disk has been extended by Fuchs 
(\cite{F04}) by embedding it into a dark halo.  All density gradients in the
halo are neglected as in the disk so that the halo density distribution is 
assumed to be homogeneous. The dark matter particles follow straight-line 
orbits with an isotropic velocity distribution modelled also by a Gaussian 
distribution. We can directly apply the results of Fuchs (\cite{F04}). 
The dark matter halo responds to the potential perturbation $\Phi_{\rm d}$ in
the disk and develops potential perturbations $\Phi_{\rm h}$ which have the 
same radial structure exp($ikx$) as in the disk. From eqns.~(26) and (28) 
of Fuchs (\cite{F04}) follows that the Fourier coefficients of the potential
perturbation in the halo at the midplane of the disk are given by
\begin{equation}
\Phi_{\rm hk} = \frac{2\pi G\rho_{\rm h}}{\sigma^2_{\rm h}}\frac{1}{k^2}
\Phi_{\rm dk}\,,
\label{eq4}
\end{equation}
where $\rho_{\rm h}$ and $\sigma_{\rm h}$ denote the density of the dark halo 
and the velocity dispersion of the dark matter particles, respectively.

This induced perturbation of the gravitational potential of the dark halo has
to be taken into account on the rhs of eq.~(\ref{eq2}),
\begin{equation}
\Phi_{\rm k} \rightarrow \Phi_{\rm dk} + \Phi_{\rm hk} \propto \Phi_{\rm dk}\,,
\label{eq5}
\end{equation}
which means that the halo supports the perturbation of the disk and the 
density perturbation in the disk is stronger than in an isolated disk.
Combining eqns.~(\ref{eq2}) to (\ref{eq5}) leads to an implicit equation that
describes the line of neutrally stable ($\omega=0$) perturbations in the space 
spanned by $Q_{\rm T}$ and $\lambda /\lambda_{\rm crit}$ as in the case of
an isolated disk, but now modified by the extra term given in eqns.~(\ref{eq4})
and (\ref{eq5}). This can be cast into dimensionless form as
\begin{eqnarray}
\alpha Q_{\rm T}^2 = \left[ 1 - {\rm exp}\left(-\frac{\alpha Q_{\rm T}^2}
{(\lambda /\lambda_{\rm crit})^2}\right)\cdot 
{\rm I}_0\left(\frac{\alpha Q_{\rm T}^2}
{(\lambda /\lambda_{\rm crit})^2} \right) \right] \frac{\lambda}
{\lambda_{\rm crit}}\nonumber \\
\times \left( 1 + \beta \, \left( \frac{\lambda}{\lambda_{\rm crit}} \right)^2
 \right)
\label{eq6}
\end{eqnarray}
with the parameters $\alpha = (3.36/2\pi )^2 = 0.286$ and
$\beta = (2\pi G\rho_h /\sigma^2_h)\cdot(2\pi G \Sigma_d)^2 /\kappa^4$.

\begin{figure}
\centering
\resizebox{\hsize}{!}{\includegraphics{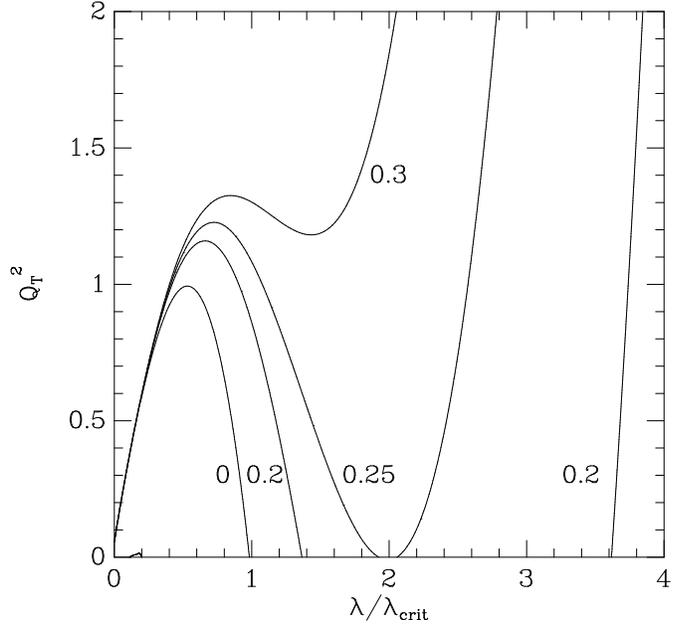}}
\caption{Separation of stable from unstable perturbations of a 
self-gravitating disk embedded in a live dark halo. $Q_{\rm T}$ denotes 
the usual Toomre stability index and $\lambda$ is the wavelength of the 
perturbation measured in units of $\lambda_{\rm crit}$. Unstable perturbations
are located in the parameter space below the dividing lines. Lines are shown 
for values of the $\beta$ parameter, which describes the dynamical
responsiveness of the dark halo, $\beta =$ 0, 0.2, 0.25, and 0.3, respectively.}
\label{fig1}
\end{figure}

In Fig.~\ref{fig1} we illustrate solutions of eq.~(\ref{eq6}) for various
values of $\beta$. The case $\beta = 0$ reproduces Toomre's classical 
(\cite{T64}) result. The unstable perturbations ($\omega^2 < 0$) are located 
in the parameter space below the line. Thus for $Q_{\rm T} \geq 1$ all 
perturbations are neutrally stable ($\omega^2 \geq 0$). This is no longer the 
case, if finite values of $\beta$ are considered. The graphs of the solutions 
shown in Fig.~\ref{fig1} always turn upwards at large wavelengths $\lambda$.
Thus at large enough wavelengths all perturbations of the disk -- halo system 
become unstable and grow exponentially. This behaviour is related to the Jeans 
collapse of the halo component. Its Jeans length is given by
 $\lambda_{\rm J} = \pi \sigma_{\rm h} / G\rho_{\rm h}$ or
\begin{equation}
\frac{\lambda_{\rm J}}{\lambda_{\rm crit}} = \frac{1}{\sqrt{2\beta}}\,.
\label{eq7}
\end{equation}
In real haloes the Jeans length will be of the order of the size of the halo
or even larger, because otherwise the haloes would have collapsed to smaller 
sizes. Thus the inferred instability of the disk -- halo system on large scales 
seems not to occur in real galaxies. As can be seen from Fig.~1 stability on 
small scales can be ensured by $Q_{\rm T}$ indices at thresholds which are 
slightly larger than in isolated disks. 

\section{Discussion and Conclusions}

As a first application of the stability criterion derived here we test the 
stability of the Milky Way disk and the surrounding dark halo in the 
vicinity of the Sun. The local disk and halo parameters listed in 
Table \ref{table1} imply $Q_{\rm T} = 2.8$ and $\beta = 0.0057$, respectively.
If we include in our estimate the cold interstellar gas with a local surface 
density of $4 M_{\odot}$/pc$^2$ (Dame \cite{D93}) and assume a velocity 
dispersion of the interstellar gas of $\sigma_{\rm g}$ = 5 km/s, which leads to 
a reduced mass weighted effective velocity dispersion of the combined stellar 
and gaseous disks, the parameter values change to $Q_{\rm T} = 2.2$ and
$\beta = 0.0078$, respectively. Equation (\ref{eq7}) implies that 
$\lambda_{\rm J}= 8 \,\lambda_{\rm crit}$ = 39 kpc. Thus the Milky Way disk 
and halo system seems to be very stable.

\begin{table}
\caption{Local parameters of the Milky Way}
\label{table1}
\centering
\begin{tabular}{ccc}
\hline
$\Sigma_{\rm d}$  & 38 $M_{\odot}/pc^{2}$ & (Holmberg \& Flynn \cite{HF04})\\
$\sigma_{\rm d}$  & 40 km/s & (Jahrei{\ss} \& Wielen \cite{Jawi97})\\
$\kappa$  & $\sqrt{2}\cdot220$ km/s/8.5 kpc & (flat rotation curve)\\
$\rho_{\rm h}$  & 0.01 $M_{\odot}/pc^{3}$ &  (Bahcall \& Soneira \cite{BS80})\\
$\sigma_{\rm h}$   &  220 km/s/$\sqrt{2}$ & (isothermal sphere)\\
$\lambda_{\rm crit}$ & 4.8 kpc & \\
$\lambda_{\rm J}$ & 39 kpc & ($\beta$ = 0.0078)\\
\hline
\end{tabular}
\end{table}

In order to explore in what range the $\beta$-parameter of spiral galaxies
is to be expected, we consider the model of a Mestel disk with the surface
density $\Sigma_d = \Sigma_0 \,R^{-1}$ embedded in a singular isothermal sphere 
representing the dark halo with the density distribution 
$\rho_h = \rho_0 \,R^{-2}$. The rotation curve of the model galaxy is given by
\begin{equation}
\upsilon^2_{\rm c}(R) = \upsilon^2_{\rm d}(R) + \upsilon^2_{\rm h}(R)
\label{eq8}
\end{equation}
with the disk contribution $\upsilon^2_{\rm d}(R)=2\pi G \Sigma_{\rm d} R$ 
= const.~and the halo contribution 
$\upsilon^2_{\rm h}(R)=4\pi G \rho_{\rm h} R^2$ = const.~(Binney \& Tremaine 
\cite{BT87}). From the radial Jeans equation follows immediately that the 
velocity dispersion of the dark matter particles is given by
\begin{equation}
\sigma^2_{\rm h} = \frac{1}{2}(\upsilon^2_{\rm d} + \upsilon^2_{\rm h})\,,
\label{eq9}
\end{equation}
because the particles are bound by both the gravitational disk and halo 
potentials. We find then
\begin{equation}
\beta = \frac{\upsilon^2_{\rm h}}{R^2}\frac{1}
{\upsilon^2_{\rm d} + \upsilon^2_{\rm h}}\frac{R^2}{4}
\frac{\upsilon^4_{\rm d}}{(\upsilon^2_{\rm d} + \upsilon^2_{\rm h})^2} = 
\frac{1}{4}\frac{\upsilon^2_{\rm h}\upsilon^4_{\rm d}}
{(\upsilon^2_{\rm d} + \upsilon^2_{\rm h})^3}\,,
\label{eq10}
\end{equation}
which implies the maximal value
\begin{equation}
\beta \leq \beta_{\rm max} (\upsilon^2_{\rm d} = 2\upsilon^2_{\rm h})
 = 0.037\,.
\label{eq11}
\end{equation}
This means that in realistic halo models its density cannot be increased,
on one hand, and the velocity dispersion of the halo particles lowered, on the 
other hand, indiscriminately, because the halo model has to stay in radial 
hydrostatic equilibrium. Equation (\ref{eq11}) implies $\lambda_{\rm J} =
3.7 \,\lambda_{\rm crit}$.  In order to ensure stability at smaller wave
lengths the Toomre stability index must be larger than $Q_{\rm T} \geq 1.02$.
We conclude from this discussion that embedded galactic disks are not prone to
Jeans instabilities, provided their Toomre stability index is a few percent
higher than $Q_{\rm T}$ = 1. From a practical point of view
the destabilizing effect of the surrounding dark halo on the Jeans instability 
of the embedded galactic disks seems to be negligible.
\begin{acknowledgements}
O.E.~gratefully acknowledges financial support by the 
International-Max-Planck-Research-School for Astronomy
and Cosmic Physics at the University of Heidelberg.
\end{acknowledgements}

\end{document}